# Time-resolved resonant elastic soft X-ray scattering at Pohang Accelerator Laboratory X-ray Free Electron Laser

Hoyoung Jang,[a)] Hyeong-Do Kim, Minseok Kim, Sang Han Park, Soonnam Kwon, Ju Yeop Lee, Sang-Youn Park, Gisu Park, Seonghan Kim, HyoJung Hyun, Sunmin Hwang, Chae-Soon Lee, Chae-Yong Lim, Wonup Gang, Myeongjin Kim, Seongbeom Heo, Jinhong Kim, Gigun Jung, Seungnam Kim, Jaeku Park, Jihwa Kim, Hocheol Shin, Jaehun Park, Tae-Yeong Koo, Hyun-Joon Shin, Hoon Heo, Changbum Kim, Changi-Ki Min, Jang-Hui Han, Heung-Sik Kang, Heung-Soo Lee, Kyung Sook Kim, Intae Eom, and Seungyu Rah

PAL-XFEL, Pohang Accelerator Laboratory, Pohang, Gyeongbuk 37673, Republic of Korea

[a)]Author to whom correspondence should be addressed: h.jang@postech.ac.kr

## ABSTRACT

Resonant elastic X-ray scattering has been widely employed for exploring complex electronic ordering phenomena, like charge, spin, and orbital order, in particular in strongly correlated electronic systems. In addition, recent developments of pump–probe X-ray scattering allow us to expand the investigation of the temporal dynamics of such orders. Here, we introduce a new time-resolved Resonant Soft X-ray Scattering (tr-RSXS) endstation developed at the Pohang Accelerator Laboratory X-ray Free Electron Laser (PAL-XFEL). This endstation has an optical laser (wavelength of 800 nm plus harmonics) as the pump source. Based on the commissioning results, the tr-RSXS at PAL-XFEL can deliver a soft X-ray probe (400–1300 eV) with a time resolution ~100 fs without jitter correction. As an example, the temporal dynamics of a charge density wave on a high-temperature cuprate superconductor is demonstrated.

## I. INTRODUCTION

In general, X-rays can be used in three main ways: spectroscopy, scattering, and imaging. Some methods, such as tomography, photoelectron microscopy, and coherent diffraction imaging, combine two or more of these techniques. When the X-ray photon energy ($E_{ph}$) is tuned at the atomic absorption edge, the electrons in the core level are excited to the unoccupied (valence) states near the Fermi level, i.e., having resonant process. Because resonance as function of $E_{ph}$ gives the footprint of the unoccupied states, which are sensitive to all configurations of degrees of freedom such as charge, spin, and orbital state of atomic sites,[1] this process is utilized in resonant X-ray techniques. Thus, using scattering at absorption edges enhances the contrast between sites and can unveil hidden periodicity, so that resonant X-ray scattering can be used to investigate electronic and spatial properties through spectroscopy and scattering ingenuity and has been a unique probe of charge, spin, and orbital order.[2] Since these degrees of freedom are important aspects of strongly correlated electronics systems and magnetic materials,[3,4] this approach can indeed contribute useful data to improve our fundamental understanding of such materials.

The energy range of soft X-rays (SX) ($\sim 10^2 < E_{ph} < \sim 10^3$ eV) is appropriate for research on materials that consist of light elements (B, C, N, and O), 3*d* transition metals, and 4*f* rare-earth elements. Thus, SX have been widely employed in spectroscopy. Despite their limitations compared to hard X-rays ($E_{ph} > \sim 4$ keV), such as longer wavelength and shorter penetration depth, resonant X-ray scattering has been demonstrated with SX.[5] Currently, resonant soft X-ray scattering (RSXS) is an important technique for exploring strongly correlated materials.[6,7] From an instrumental perspective, dedicated RSXS instruments are considered to be essential in synchrotron facilities worldwide.[7]

Nowadays, with the increased scientific interest in compounds containing 3*d* transition metals and 4*f* rare-earth elements that exhibit magnetism, complex ordering, and high-$T_c$ superconductivity, as well as the development of ultrahigh vacuum and machining technologies, RSXS instruments have been continually improved. In particular, brighter light sources have boosted the application of X-ray techniques even more. After the first light at the Linac Coherent Light Source,[8,9] X-ray free-electron laser (XFEL) facilities, which generate high-intensity femtosecond pulses, were constructed in several locations around the world.[10–14] For example, the SXR instrument[15–18] of LCLS has been successfully used for ultrafast SX experiments as well as time-resolved resonant soft X-ray scattering (tr-RSXS) experiments.[19–25]

Thus, the *Republic of Korea* decided to construct an XFEL facility at Pohang Accelerator Laboratory and named PAL-XFEL.[11,12] It was planned to develop a tr-RSXS system at this facility during initial construction.[11,26] Thus, here, we present the development and first commissioning results of the tr-RSXS system at the soft X-ray scattering and spectroscopy (SSS) beamline[26] of PAL-XFEL. Section II briefly introduces the SSS beamline and operational configuration, which is in the user operation from 2017. In Section III, we introduce the tr-RSXS instrumentation. In Section IV, based on the commissioning results, we evaluate the performance of the tr-RSXS system at PAL-XFEL. Finally, in Sections V and VI, we give an outlook and summary.

## II. SSS BEAMLINE AT PAL-XFEL

The SSS beamline was uniquely designed and developed to cover the SX range (250–1300 eV) at PAL-XFEL.[26] As a first step, the beamline has been utilized for regular user programs since 2017, mainly X-ray absorption spectroscopy (XAS) and X-ray emission spectroscopy (XES) experiments.[27,28] In the second stage, in which the capability of the beamline was expanded, the RSXS endstation was integrated with the SSS beamline. From 2020, the tr-RSXS system has been considered as a general user program at PAL-XFEL. Since the previous commissioning report,[26] which thoroughly described the SSS beamline and optics, here we simply revisit the overall beamline configuration and specification and describe some changes.

To produce SX at PAL-XFEL, an electron beam is accelerated up to 3.15 GeV and travels to 35-mm-period planar undulators.[29] This generates a free-electron laser beam through self-amplified spontaneous emission (SASE).[9,30] This so-called pink beam has a ~0.5% (FWHM) bandwidth around the central $E_{ph}$.[26] At 850 eV, the pink-beam pulse energy is estimated to be about 200 μJ per pulse, which corresponds to ~2 × $10^{12}$ photons per pulse.[26] While maintaining the electron beam energy at 3.15 GeV, the incident pink beam with $E_{ph}$ between 400 eV and 1300 eV is conveniently tunable by changing the size of the undulator gap. Note that such $E_{ph}$ range is generally supported at the RSXS endstation and the photon number per pulse varies from 5 × $10^{12}$ (400 eV) to 1 × $10^{12}$ (1300 eV) which is estimated by electron energy loss measurement and mirror reflectivity correction. Comparing the X-ray pulse intensities measured by photodiode and considering the pink beam photon number, the photon numbers per pulse of the monochromatic beam with 100 μm exit slit width are estimated as 1.7 × $10^{10}$ at 600 eV and 2.5 × $10^{10}$ at 900 eV.

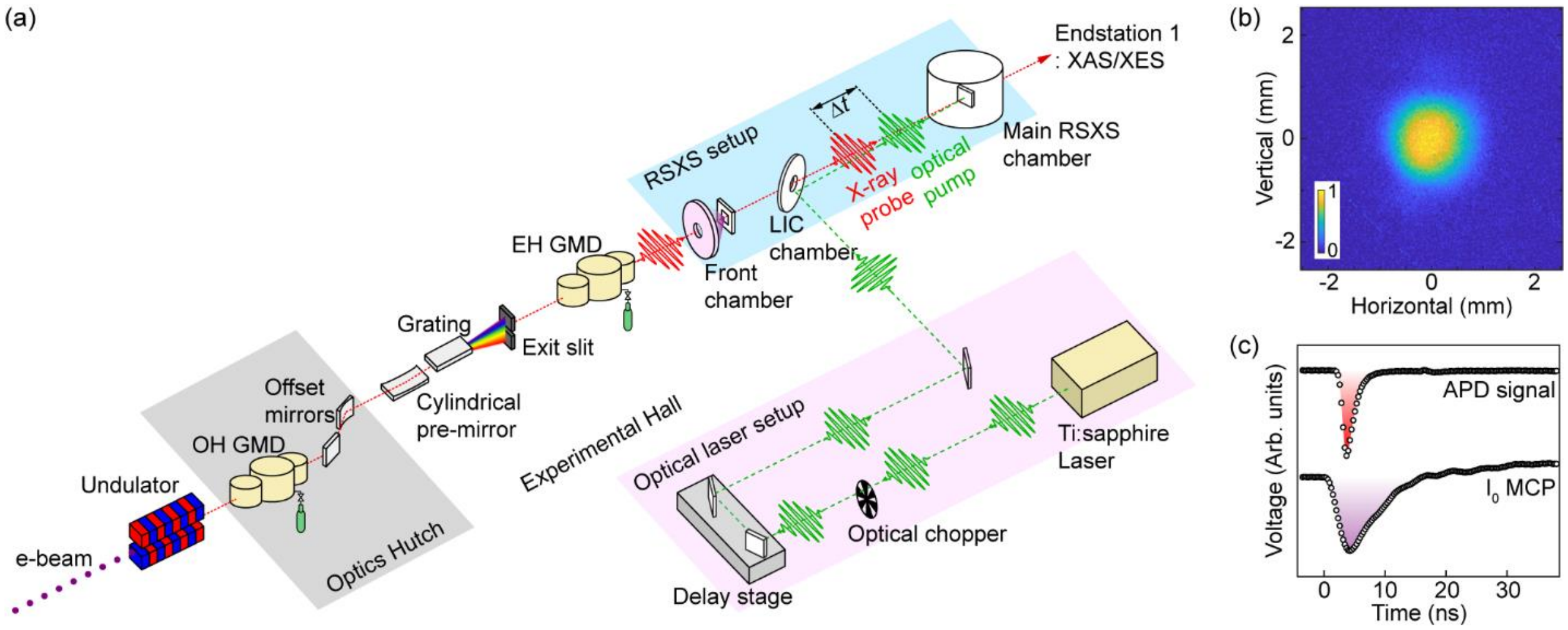

**FIG. 1.** (a) Schematic of the SSS beamline. From upstream (lower left), the figure shows (1) the accelerated electron beam and the planar undulator, (2) the OH-GMD and horizontal offset mirrors in the OH (gray), (3) the pre-mirror and grating set, (4) the EH-GMD, and (5) the RSXS system (light blue). The optical laser system is shown in pink at the lower right. (b) Single shot image on a PM after the offset mirrors in the OH, measured for a 900-eV SASE free-electron laser beam. (c) RSXS APD and $I_0$ signals recorded on a high-speed digitizer.

Figure 1 schematically shows the main components of the SSS beamline. The SASE free-electron laser beam produced at the undulator travels to the optics hutch [OH, gray area in Fig. 1(a)]. Four pop-in monitors (PMs) are installed between the components in the OH to destructively diagnose the position and shape of the X-ray beam. Each PM has a disk-shaped Ce:YAG (hereafter simply YAG) crystal in the vacuum chamber, which is mounted on a vertical motion manipulator. The YAG fluorescence image is reflected by a mirror tilted at 45° and projected onto an out-of-vacuum digital camera (MANTA G-046B, Allied Vision), which is mounted on a glass window flange. Each PM can record single-pulse images up to 60 Hz with full image resolution. An example of a single-pulse image is shown in Fig. 1(b). A gas monitor detector (GMD),[31,32] upstream in the OH, monitors the incoming XFEL pulse intensities shot-to-shot. A pair of horizontal offset mirrors, downstream in the OH, reject the bremsstrahlung background as well as higher-order contamination. The second offset mirror has a cylindrical surface and its horizontal focal point is the RSXS endstation.

After passing through the OH, the X-ray beam arrives at the experimental hall (EH). Upstream in the EH, there are a pair of planar varied line-spacing gratings (100 lines/mm and 200 lines/mm) and a cylindrical pre-mirror. There are two beam modes, pink and monochromatic. In pink-beam mode, instead of a grating, a planar reflection mirror is used to ensure the maximum flux arrives at the endstations. In monochromatic-beam mode, which is expected to be used for most of experiments in the RSXS endstation, the grating fixes $E_{ph}$ and the size of the exit slit determines the bandwidth of $E_{ph}$ and the incident photon flux. The gap size is set to maximize the photon flux. Note that the horizontal size of the X-ray beam about 100 μm after focusing by horizontal focusing mirror and the vertical size at the RSXS endstation is also determined by the size of the exit slit because there are no focusing optics. In Table 1, the vertical and horizontal X-ray spot sizes at the sample position in the RSXS endstation and energy bandwidth are presented in cases of selected $E_{ph}$ (640 eV, 710 eV, 850 eV, and 930 eV which are $L_3$-edges of Mn, Fe, Ni, and Cu, respectively) and the exit slit sizes of 100 μm and 200 μm. Note that spot sizes were measured by knife-edge scanning and the energy bandwidth according to the exit slit size is calculated from grating specifications. Another GMD has recently been installed downstream of the exit slit and is being commissioned. This EH-GMD monitors either the pink-beam intensity in the EH after rejecting contamination or the monochromatic-beam intensity. It can be utilized as an $I_0$ monitor. Both the OH-GMD and EH-GMD use krypton at a pressure of $10^{-6}$ Torr level as the gas medium. The RSXS system is downstream of the EH-GMD. The XAS/XES endstation is in the very downstream part of the SSS beamline.[26,27]

| $E_{ph}$ (eV) | Slit size (μm) | Energy bandwidth (eV) | Vertical size (μm) | Horizontal size (μm) |
|---|---|---|---|---|
| 650 | 200 | 0.105 | 150 | 137 |
| | 100 | 0.053 | 83 | 101 |
| 710 | 200 | 0.152 | 186 | 118 |
| | 100 | 0.076 | 109 | 91 |
| 850 | 200 | 0.265 | 197 | 111 |
| | 100 | 0.133 | 115 | 105 |
| 930 | 200 | 0.343 | 196 | 99 |
| | 100 | 0.171 | 98 | 97 |

**TABLE 1.** Summary of the calculated energy bandwidth and measured X-ray spot sizes

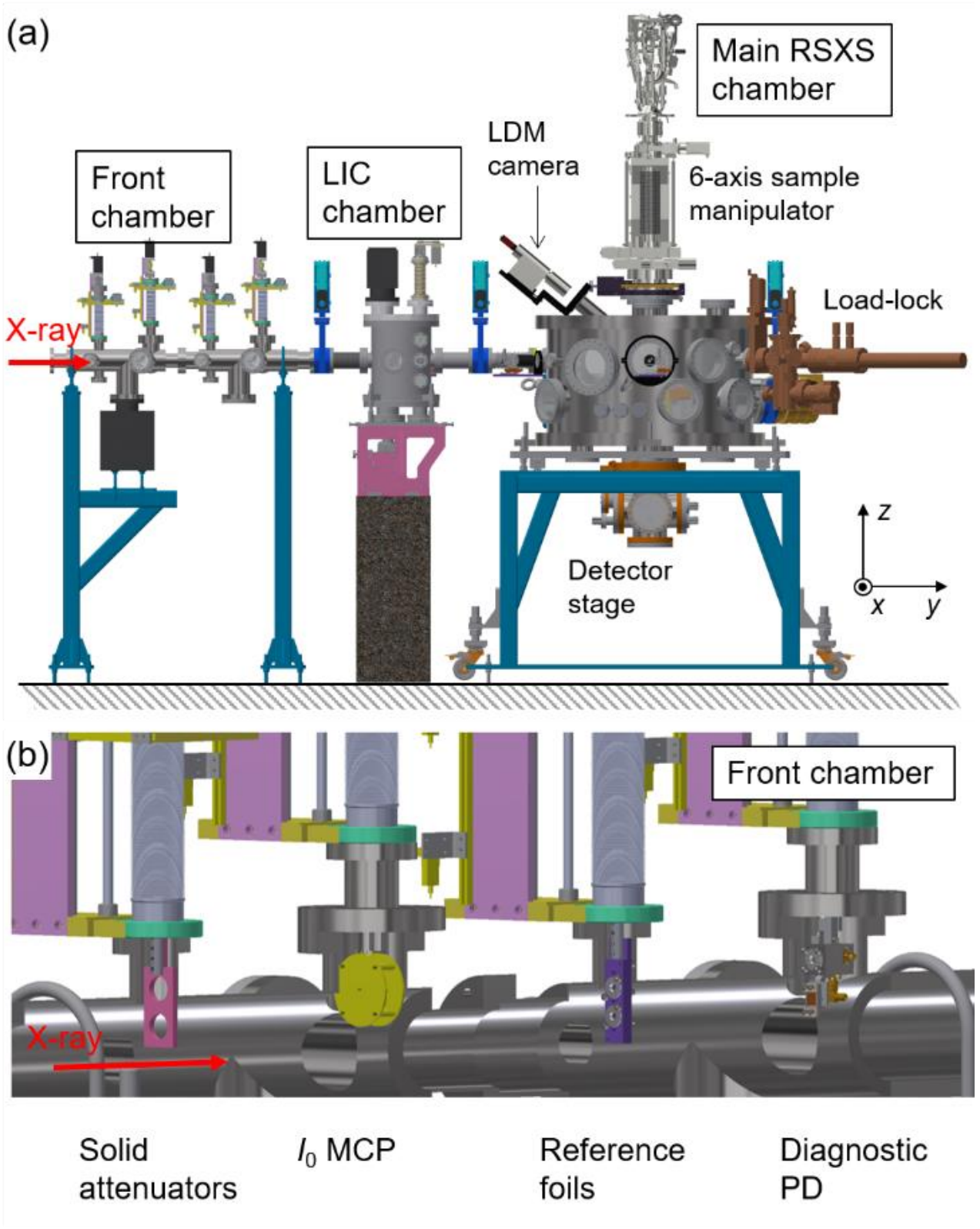

**FIG. 2.** (a) Model of the RSXS system. Front, LIC, and main chambers from upstream. (b) General setup of the front chamber: solid attenuators, $I_0$ MCP, reference foils, and diagnostic PD from upstream.

# III. TR-RSXS INSTRUMENT AT SSS BEAMLINE

## A. Overview

The RSXS system is in the middle of the SSS beamline and downstream of the EH-GMD, as shown in light blue in Fig. 1(a). It has three main chambers in series, going from upstream: the front chamber, the laser-in-coupling (LIC) chamber, and the main RSXS chamber. A detailed model is shown in Fig. 2(a). Note that these chambers are aligned at the X-ray height, which is 1417.5 mm above the floor. The front chamber is used for beam diagnostics and $E_{ph}$ calibration and has the $I_0$ monitor. The LIC chamber synchronizes the optical paths of the optical laser and X-ray beams. The main RSXS chamber houses the tr-RSXS experiments.

### *1. Front chamber*

This chamber can integrate four components and the configuration can be altered depending on the requirements. In general, it has solid attenuators, a microchannel plate (MCP) for $I_0$ monitoring, reference foils, and a diagnostic photodiode (PD), as shown in Fig. 2(b). Each component is mounted on an XZ manipulator, which can move the component perpendicular to the X-ray beam. The solid attenuators reduce the intensity of the incoming X-ray beam. They are composed of various combinations of aluminum foil of different thicknesses to cover the broad $E_{ph}$ range. The $I_0$ MCP assembly consists of a two-stage MCP with a 4-mm-diameter central hole (F2223-21SH, Hamamatsu) with its effective surface facing downstream. The X-rays, which pass through the MCP central hole, shine onto a Pt-deposited 200-nm-thick $Si_3N_4$ membrane and the backscattered intensity is measured by the MCP to normalize the signal.[18,26] An Si PIN PD (S3590-09, Hamamatsu) is used for various diagnostic purposes. A small amount of external bias (up to ~20 V) can be applied to the PD. The reference foils are 100-nm-thick 3*d* transition metal foils (Fe, Co, Ni, and Cu) with a 100-nm Parylene N support (Lebow Company). By using the MCP as an $I_0$ monitor and the PD as a signal detector, the foils can be used to calibrate $E_{ph}$ of the X-rays via transmission XAS. A high-speed digitizer records the signal from the MCP and PD on a shot-to-shot basis, the details of which are described in Section III.C.

### *2. LIC chamber*

This chamber is simply composed of a hole-mirror and a PM. The hole-mirror is a 2-in.-diameter fused silica disk mirror coated by protected aluminum with a central hole. X-rays can pass through the central hole while the optical laser beam is reflected from the surface. The optical laser beam is injected at an angle of 90° to the X-ray beam and becomes nearly parallel to it (<1°) after being reflected. The hole-mirror has four axes of manipulation. Two horizontal translations, parallel and perpendicular to the direction of X-ray propagation, are controlled by out-of-vacuum stepping motor stages connected via a bellows welded onto the bottom of the chamber. Vertical and horizontal tilting of the mirror mount is controlled by piezo-based linear actuators (N-480, Physik Instrumente). By tuning the reflection angle of the optical laser, both ultrashort pump and probe pulses can illuminate the same spot of the sample surface in the main chamber. A PM is used to position the hole-mirror and to monitor the X-ray beam.

### *3. Main RSXS chamber*

This multi-purpose vacuum chamber is a 900-mm-inner-diameter cylindrical stainless-steel vessel. It has bottom and top flanges, which can be opened to allow the equipment inside the chamber to be changed. A turbomolecular pump and an ion pump are installed and the base pressure is a level of $2 \times 10^{-8}$ Torr. The following are descriptions of the components inside the chamber.

*a. Six-axis sample manipulator.* This manipulator can move over six axes ($\theta$, $\phi$, $\chi$, $x$, $y$, and $z$) as indicated in Fig. 3. As shown in Fig. 2(a), the sample manipulator is assembled on the central port of the top flange. On the top of the port, there is a 6-in.-inner-diameter differentially pumped rotating flange (DPRF, RNN-600, Thermionics), which can be used to rotate the sample horizontally ($\theta$). There is an XYZ translator on top of the DPRF. The scientific sample slot has an in-vacuum motorized tilting mechanism, such that the rotation axes are perpendicular ($\phi$ ) and parallel ($\chi$) to the sample surface. The detailed information of six axes in the sample manipulator is summarized in Table 2.

*b. Detector stage.* There is another DPRF (RNN-800, Thermionics) on the central port of the bottom flange of the main chamber, which is used to integrate a detector stage [Fig. 2(a)]. It allows horizontal rotation of the detector ($2\theta$). The scattering signal is counted with an avalanche photodiode (APD; SAR3000x, Laser Components), which is utilized as a point detector. The APD is enclosed in an aluminum cage. Its entrance is covered by a 300-nm-thick aluminum filter to block stray light and the optical laser. The APD is directly connected to the coupling circuit, which consists of a resistor and a capacitor. The SMA signal cable (KAP-1CX-39SMA, Accu-Glass Products) is connected from the APD cage to the SMA feedthrough in the bottom of the detector stage. The voltage bias is applied from a BNC feedthrough. The APD has a 3-mm-diameter active area and the solid angle from the sample can be altered by changing the distance to the sample (80–160 mm) or using a smaller hole aperture. Up to three APDs can be used, and their configuration is determined by the experimental requirements. The APDs are mounted on a vertical goniometer (WT-120, Physik Instrumente) and can be selected during an experiment. The detailed information of two axes in the detector stage is summarized in Table 2. Moreover, two PDs are mounted together. One is a Si PIN PD for calibrating and monitoring the X-ray and optical laser intensities and the other is a fast GaAs PD (G4176, Hamamatsu) with a rise time of 30 ps, which is used to measure the coarse temporal overlap between the X-ray and optical laser beams.[33] Finally, a channel electron multiplier (KBL15RS, Sjuts) was installed to measure the fluorescence or electron yield XAS signal of a sample. All signal and bias voltage cables are connected to the electrical feedthrough flanges on the bottom of the detector stage, and they rotate with the detector stage. A model of the detector stage with a sample cryostat is shown in Fig. 3.

| **Stage** | **Axis** | **Range** | **Resolution** | **Motor** | **Encoder** | **Manufacturer and notes** |
|---|---|---|---|---|---|---|
| 6-axis sample stage | $x$ | ±10 mm | 1 μm | ST4118L1804-B | HEDL-5540-H14-400 | Nanotec, assembled by PreVac |
| | $y$ | ±10 mm | 1 μm | ST4118L1804-B | HEDL-5540-H14-400 | |
| | $z$ | 200 mm | 1 μm | AS5918L4204-E | Motor integrated rotary encoder | |
| | $\phi$ | 360° | 0.1° | AS4118L1804-E | | |
| | $\chi$ | -20 to +40° | 0.1° | AS4118L1804-E | | |
| | $\theta$ | -30 to +200° | < 0.01° | 8718M-22D-01RO | - | LIN Engineering, assembled by Thermionics |
| Detector stage | $2\theta$ | ±180° | < 0.01° | 8718M-22D-01RO | - | |
| | $\gamma$ (flip) | ±40° | 0.01° | in-vacuum 2-phase stepper motor | - | Assembled by Physik Instrumente |

**TABLE 2.** Summary of motorized axes.

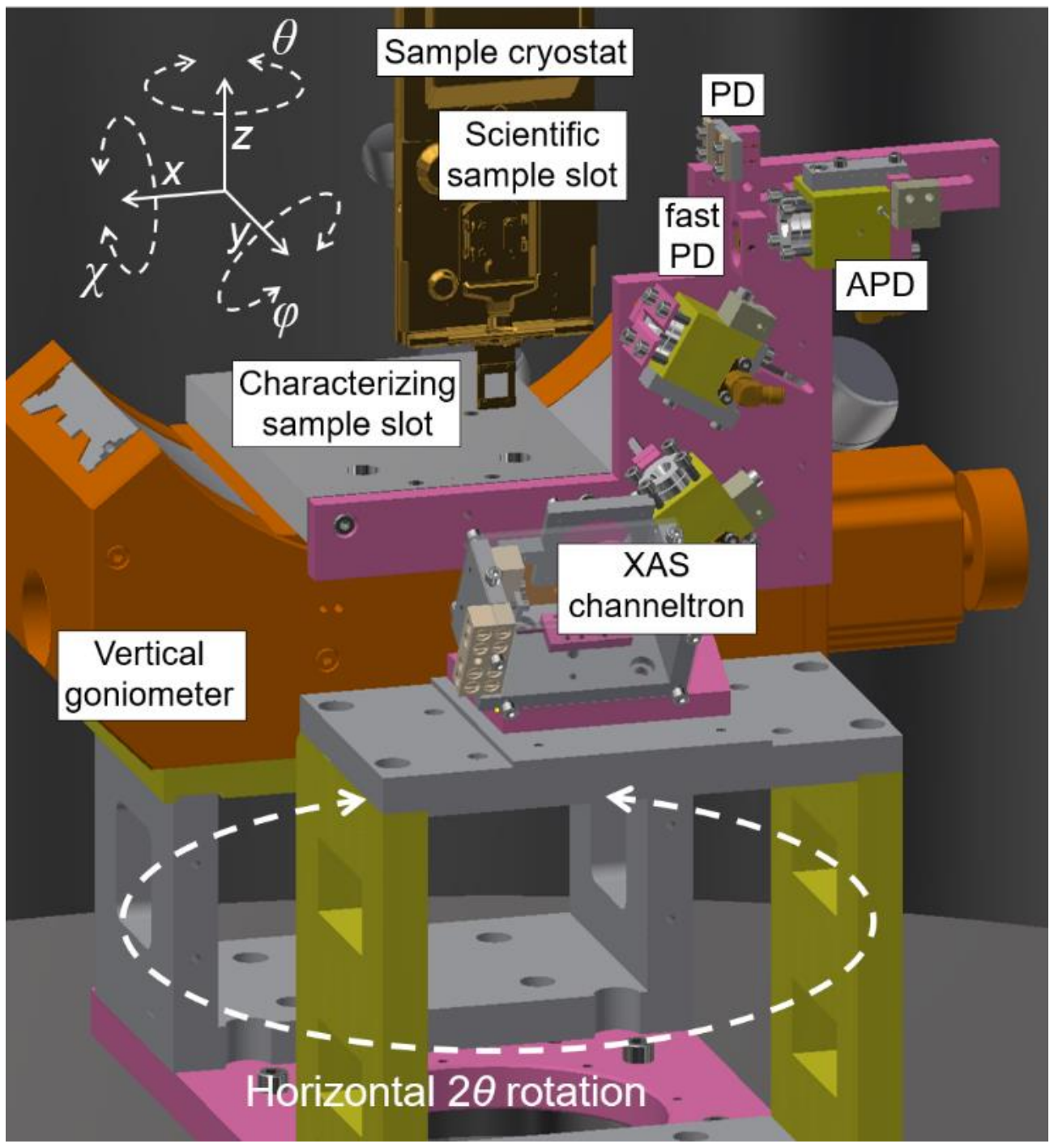

**FIG. 3.** Model of detector stage and sample cryostat. 6 axes of the sample manipulator and $2\theta$ rotation direction are presented.

*c. Error of pointing of rotating axes.* We modified a conventional six-axis cryostat manipulator (PreVac), which was manufactured for an angle-resolved photoemission experiment. Moreover, the sample ($\theta$) and detector ($2\theta$) rotation axes are provided by separated DPRFs. Therefore the error of pointing, i.e., sphere of confusion, of this setup is expected to be large. The difference between the $\theta$ and $2\theta$ rotation axes is about 500 μm. Note that the DPRFs were mounted as parallel as possible (<0.1°) using a level meter. Because *x*- and *y*-axes translator is on top of $\theta$-rotation DPRF, the $\phi$ (azimuth) and $\theta$ rotation axes can well match if the sample is properly mounted. Though the flip ($\chi$) rotation axis is close to the $\phi$ and $\theta$ rotation axes, the sample surface is 8 mm away from the $\chi$ rotation axis to secure X-ray path to the detector when $\theta$ is 0°. Therefore, the sample position should be properly tuned by translating *y*- and *z*-axes to ensure the X-ray beam spot on the same sample position while $\chi$ is rotating. The $\gamma$ (detector flip) rotating axis is about 1 mm off from the $2\theta$ rotation axis.

*d. Scientific sample slot and temperature control.* The holder for a scientific sample is a conventional flag-style sample holder, which can be transferred from the load-lock chamber using a magnetic transfer arm [brown color in Fig. 2(a)]. In the load-lock chamber, up to eight sample holders can be mounted. In general use, a sample crystal or film is glued onto an adapter and the adapter is attached to the sample holder. The temperature of the sample is controlled by an open-cycle liquid helium/nitrogen flow and an in-vacuum heater. There is a closeable radiation shield around the scientific sample slot, which allows it to be cooled down to ~15 K. The highest temperature is 400 K.

*e. Characterizing sample slot.* On the bottom of the radiation shield, there is an additional sample slot without tilting motion, on which characterizing samples can be mounted. We mounted a YAG crystal in this slot to assess the spatial and temporal overlap of the laser and X-ray pulses.

*f. Long-working distance microscope camera.* There is a port on the top flange tilted at 40° in the upstream direction. A long-working distance microscope (LDM; UWZ-500F, Union Optics) with a working distance of 500 mm and a digital camera (MANTA G-046B, Allied Vision) are mounted on this port [Fig. 2(a)]. This LDM camera is used to measure the spatial and temporal overlap between the X-ray and optical laser pulses as well as for sample alignment.

*g. Other cameras and sample alignment.* In addition to the LDM camera, additional digital cameras (MAKO G-507B, Allied Vision) are installed on the glass window flanges at X-ray height on the main chamber to observe the sample inside the chamber. Using the LDM camera and a digital camera located in 90° to the X-ray direction and comparing their images at $\theta = 0°$, 90°, and 180°, the sample surface can be put on the rotation axis by translating *x*- and *y*-axes. After this alignment, rotation axis is confirmed by 'half-cut' procedure. If the X-ray beam does not exactly pass the rotation axis, the main chamber can be translated perpendicular to the X-ray direction.

## B. Optical laser system

A 4-mJ Ti:sapphire laser system has been installed in the SSS beamline. It can deliver an optical laser beam for the XAS/XES and RSXS endstations.[33] The fundamental wavelength of the optical laser has a duration of 40 fs. The maximum repetition rate is 60 Hz in typical pump–probe experiments while the laser operates at up to 1080 Hz. As described in Section III.A.2, the optical laser and X-ray beams are spatially synchronized in the LIC chamber. Currently at the RSXS endstation, a fundamental 800-nm-wavelength optical laser beam (1.55 eV) and its harmonics at 400 nm and 266 nm are available. The optical laser energy is controlled by a motorized attenuator consisting of a half-wave plate and two broadband thin-film polarizers (Watt Pilot, Altechna). The delay time ($\Delta t$) between the X-ray beam and the optical laser is controlled by a mechanical delay stage (IMS400LM, Newport). The maximum ±200 mm travel distance enables a delay time of up to 2.6 ns. A beam stabilizer system (Compact, MRC Systems) mitigates the pointing jitter on the sample, which can arise during delivery of the laser beam. The laser spot size is measured by a laser beam profiler which locates in the outside of the chamber, but the laser path length to the beam profiler is same to one to the sample in main chamber. Moreover, the laser spot size is also confirmed by knife-edge scanning. The details of the optical laser system and how it is synchronized with PAL-XFEL are described in Kim *et al.*[33]

## C. Instrument control and data acquisition

The PAL-XFEL facility has officially adopted the Experimental Physics and Industrial Control System (EPICS)[34,35] for the entire control system. EPICS enables convenient remote access and control via process variables that are inherently assigned to devices. The SSS beamline and RSXS endstation are also mostly controlled by EPICS.

We have developed Python-based software, named PXC (PAL-XFEL controller), for RSXS experiments. The software is compatible with SPEC software (Certified Scientific Software), which has been employed globally for X-ray scattering experiments. Control of optical laser and X-ray parameters and manipulation of the sample are integrated in PXC and counts from detection devices, e.g., $I_0$ MCP, PD, and APD, are recorded for a given number of X-ray pulses at each step of an experiment. PXC is continually being updated.

The input signals are processed by a four-channel 12-bit high-speed digitizer (ADQ412DC, Teledyne SP Devices), which has an input bandwidth of 925 MHz and a sampling rate of $2 \times 10^9$ samples per second. Figure 1(c) displays an example of raw data from the APD and $I_0$ MCP. Signals from the OH-GMD and EH-GMD can be simultaneously processed with separate 10-bit digitizers (PXIe-5160, National Instrument), which have an input bandwidth of 500 MHz and a sampling rate of $1.25 \times 10^9$ samples per second. When the input signal is weak, a current amplifier (HCA-400M-5K-C, FEMTO: 5kV/A gain with 400 MHz bandwidth) can be employed.

During data acquisition, raw data are recorded from all channels and the integrated counts of a specific region of interest are also saved for convenience. Since all pulses have a pulse ID and timestamp, a shot-to-shot analysis is possible. The data are stored in standard HDF5 format and analyzed by software coded in Matlab or Python based on users' preference.

## IV. TR-RSXS COMMISSIONING RESULTS

### A. X-ray and optical laser cross-correlation

In an optical laser pump and X-ray probe experiment, two of the most important of the experimental parameters are the spatial and temporal overlaps between the two pulses at the sample position. As widely used,[14,26,36,37] we utilize a YAG crystal to assess both the spatial and temporal overlaps. The LDM camera mounted on the main RSXS chamber observes the YAG crystal in the characterizing sample slot. Note that the high-intensity pink beam is used to assess the cross-correlation of the X-ray pump and optical laser probe.

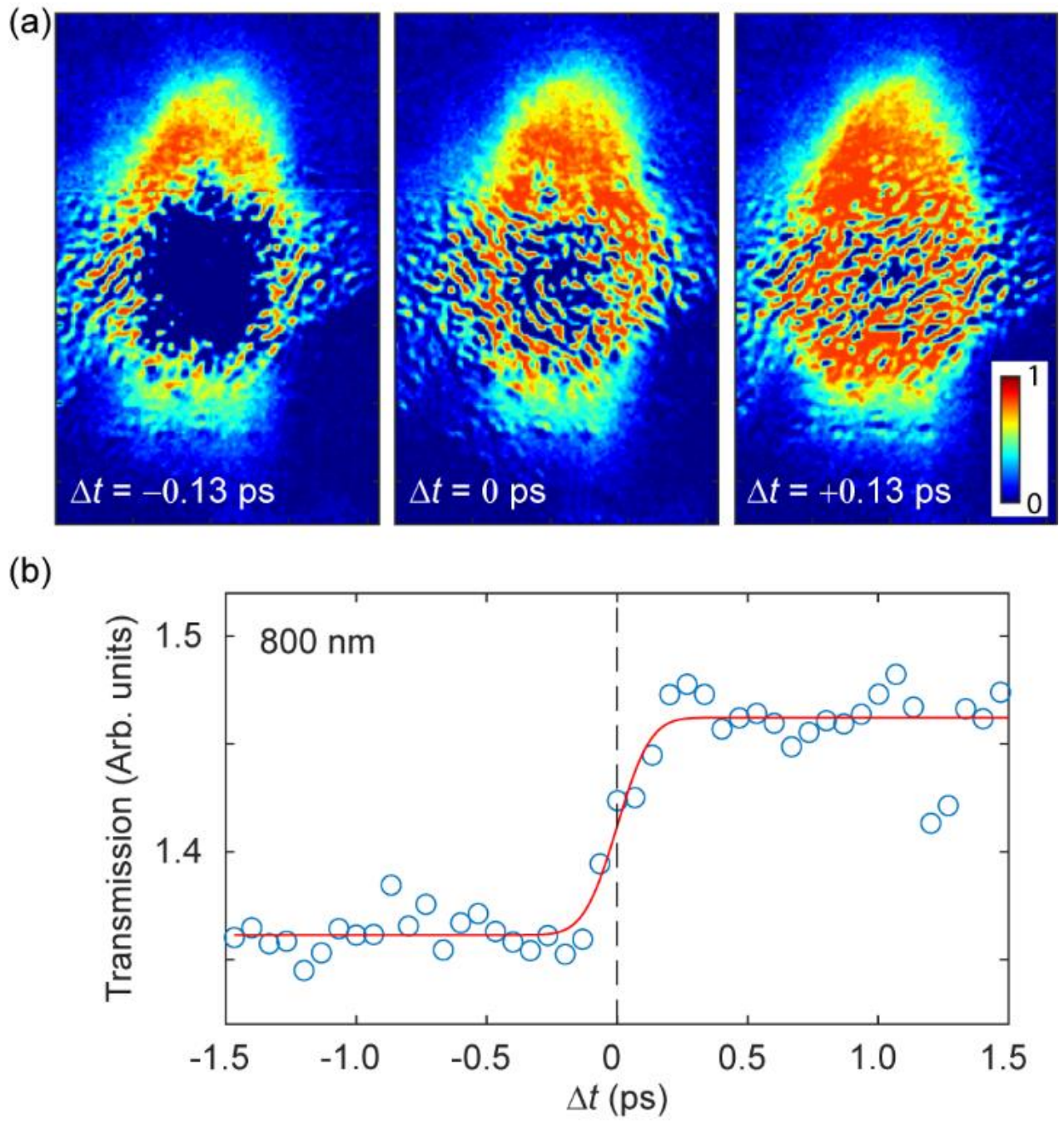

**FIG. 4.** Optical laser and X-ray cross-correlation. (a) Difference images of an optical laser spot diffusively reflected from the Ce:YAG crystal holder, as observed by the LDM camera. The delay time was -0.13 ps (left), 0 ps (middle), and +0.13 ps (right). $E_{ph}$ = 900 eV in pink-beam mode. The wavelength of the optical laser was 800 nm. (b) Optical laser transmission as a function of delay time ($\Delta t$). The solid line is the error function fit.

The spatial overlap is first checked by comparing the spot positions of the X-ray and optical laser beams. If necessary, the position of the laser spot is adjusted by tilting stages, either in the LIC chamber or the laser system. Since a YAG crystal does not fluoresce when illuminated by an 800-nm optical laser, we utilize the transmitted intensity after the beam has passed through the crystal and is diffusively reflected back from the YAG holder.

To assess the temporal overlap, the delay time of the optical laser was adjusted by moving the mechanical delay stage while the spots overlapped spatially on the YAG crystal. When an X-ray shines on a YAG crystal and is absorbed, the photoionization and subsequent cascade ionization increase the free-carrier density, which results in a rapid change in the optical properties of the YAG crystal, and therefore, the transmission of the optical laser through the crystal changes.[36,38] These changes depend on the difference in the arrival times of the two pulses, i.e., optical laser and X-ray cross-correlation, and so we can find the time zero ($t_0$) at the RSXS endstation. To obtain clearer images, the optical laser repetition rate was set to twice the X-ray repetition rate, i.e., 60 Hz for the optical laser and 30 Hz for the X-rays. The optical laser spots after transmission through the YAG crystal were recorded with and without an X-ray pulse. When the X-ray arrives after the optical laser ($\Delta t > 0$), the fluorescence due to the X-ray pulse is mainly visible [Fig. 4(a), right panel]. In contrast, when the X-ray arrives before the optical laser ($\Delta t < 0$), the optical laser transmission and corresponding diffuse scattering from the holder become weaker, resulting in a decrease in the intensity in the area where the two pulses overlap spatially [Fig. 4(a), left panel]. The spatial overlap between two pulses can also be seen in this image.

Moreover, the optical laser transmission can be recorded directly by the PD inside the main chamber [Fig. 4(b)]. The cross-correlation data were fitted by a function [24,38]:

$$I = I_0 + \frac{I_1 - I_0}{2}\left[1 + \mathrm{erf}\left(\frac{\Delta t - t_0}{\sqrt{2}\tau_{\mathrm{res}}}\right)\right] \quad (1)$$

where erf is the error function, $t_0$ is the center of the error function and the estimated time zero, $\tau_{\mathrm{res}}$ is the time resolution, and $I_0$ and $I_1$ are the transmitted intensities before and after the X-ray pump, respectively. Using this fitting, we estimated that $\tau_{\mathrm{res}} \approx 110$ fs without jitter correction. The cross-correlation shown here used an 800-nm-wavelength optical laser and a 900-eV X-ray beam.

## B. tr-RSXS demonstration

As a part of the tr-RSXS commissioning, we employed a high-$T_c$ cuprate superconductor. This is one of the most studied materials in condensed matter research, especially by XFEL facilities[39–41] and using SX scattering.[23–25] Thus, we investigated the archetypal Y-based cuprate, single-crystal $YBa_2Cu_3O_{6.73}$ (YBCO).[42–44] $YBa_2Cu_3O_{6+x}$ can have a charge-density-wave (CDW) order, which shows complex intertwining phenomena with superconductivity.[25,40,41,43–45] In this context, exploring the CDW order of the high-$T_c$ cuprate would be a good option for tr-RSXS commissioning. Note that the $YBa_2Cu_3O_{6+x}$ system contains an oxygen order (OO) in the charge reservoir layer.[47] In current doping with $x \approx 0.73$, the OO is ortho-III type, resulting in an OO with an in-plane wavevector ~1/3.[43,44]

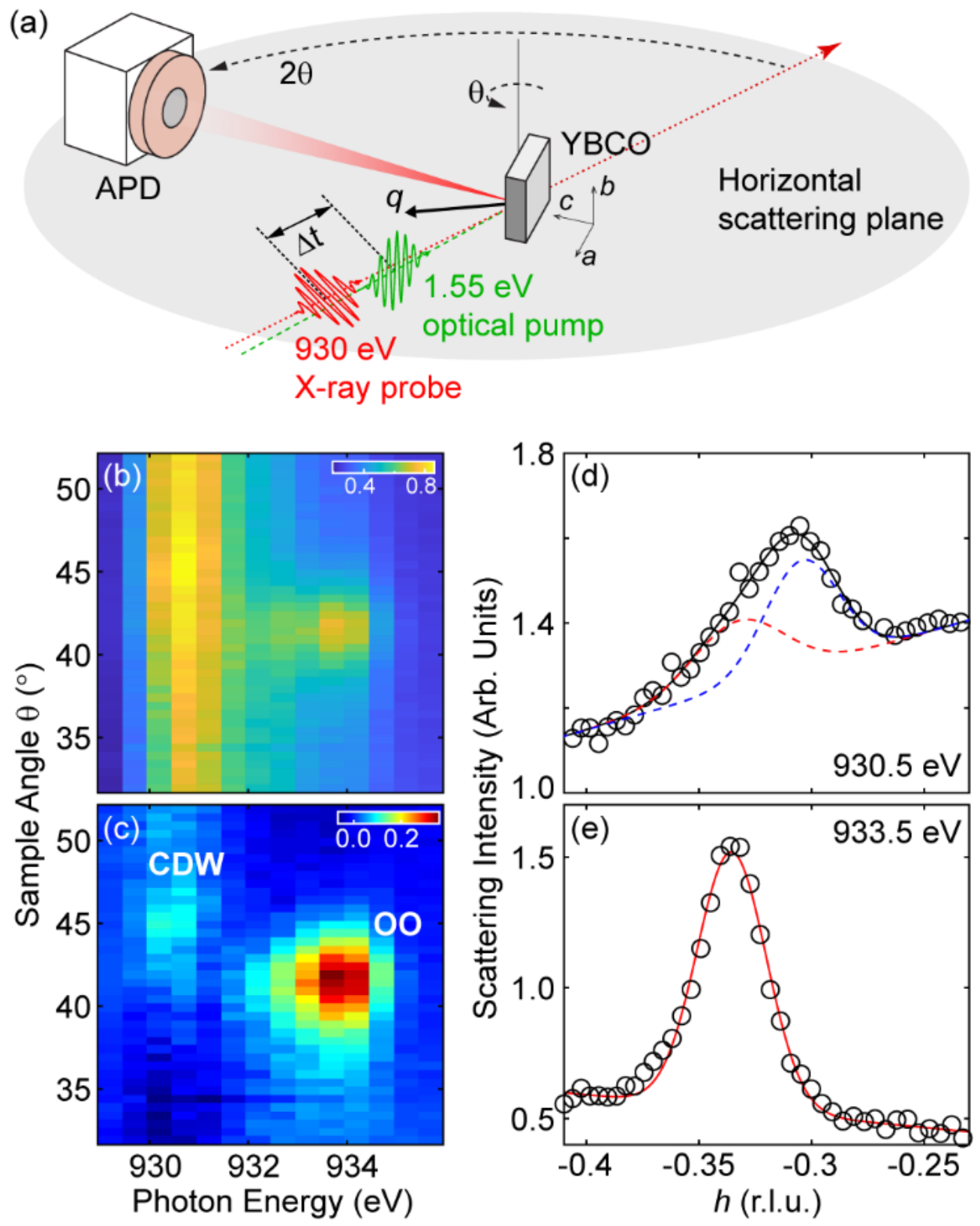

**FIG. 5**. (a) Experimental geometry of tr-RSXS experiment with YBCO. (b) Map of intensity for $\theta$ vs $E_{ph}$ of YBCO RSXS at Cu $L_3$-edge at $T$ = 70 K. (c) Map of intensity for $\theta$ vs $E_{ph}$ after subtraction of the fluorescence background. The CDW and OO peaks are indicated. $h$ profile at (d) 930.5 eV and (e) 933.5 eV. In (d) and (e), the peaks are, respectively, fitted by one and two Gaussian curves with a linear slope background. The red and blue curves are for the OO and CDW, respectively.

We first carried out a static RSXS experiment without an optical pump. During this experiment, we used the monochromatic-beam mode. The exit slit was 200 μm wide. The incident X-rays illuminated an area on the wide sample surface, which was perpendicular to the crystalline $c$-axis. The scattering plane was parallel to the crystalline $ac$-plane [Fig. 5(a)]. Due to the 2-dimensional (2D) nature of CDW and OO,[43–46] it is possible to explore the $h$-space by rotating the sample through angle $\theta$ and the APD detector through angle $2\theta$. According to previous work on $YBa_2Cu_3O_{6+x}$, the strength of the CDW order is maximized at the superconducting critical temperature ($T_c$).[40,41,44,45] In this context, we set the sample temperature ~70 K, which is the $T_c$ of YBCO.

Figure 5(b) presents the Cu $L_3$-edge RSXS intensity map of YBCO as a function of incident $E_{ph}$ and sample angle $\theta$ with $2\theta$ fixed at 156°. The measured $\theta$ range (from 32° to 52°) corresponds to $l \approx 1.4$ in ($h$ 0 $l$) reciprocal space, which covers $h \approx -0.31$, which is expected for the CDW wavevector. We found quasi-2D intensities at $E_{ph} \approx 930.5$ eV and $\theta \approx 45°$ ($h \approx -0.31$) and at $E_{ph} \approx 933.5$ eV and $\theta \approx 42°$ ($h \approx -0.33$), indicating CDW and OO, respectively. Both intensities are masked by the strong Cu fluorescence background.[43] Thus, we subtracted the fluorescence background signal estimated as the average signal for the highest-angle (52°) and lowest-angle (32°) regions where there are no ordering features. As shown in Fig. 5(c), the two peaks are clearly visible after subtracting the fluorescence background. Figures 5(d) and 5(e) show the $h$ profiles for 930.5 eV and 933.5 eV, which, respectively, correspond to CDW and OO. Note that $2\theta$ was fixed at 166° for this $h$ profile measurement. The peak profile at 930.5 eV also shows the OO tail at $h \approx -0.33$. Therefore, we fitted the $h$ profile measured at 930.5 eV using two Gaussian curves with a linear slope background [Fig. 5(d)]. The blue dashed line near $h \approx -0.31$ and the red dashed line near $h \approx -0.33$ represent the CDW peak and OO tail, respectively. On the other hand, the $h$ profile measured at 933.5 eV was fitted by a single Gaussian with a linear slope background. There is clearly a strong OO peak at $h \approx -0.33$ [Fig. 5(e)]. The data in Fig. 5 are qualitatively consistent with the previous results which were measured in the synchrotron facilities.[43,44] Note that each theta scan in Figs. 5(d) and (e) took about 100 seconds to measure a full curve with 60 Hz X-ray repetition rate.

Since we have explored the static features sufficiently, we move to a temporal dynamics study of the CDW order. Note that we adopt the same experimental conditions, namely the sample temperature ($T = 70$ K), detector angle ($2\theta = 166°$), and $E_{ph}$ (930.5 eV), in this temporal dynamics experiment. For the tr-RSXS experiment with YBCO, the wavelength of the optical pump was set to 800 nm (1.55 eV). $\Delta t$ was controlled by the delay stage after checking the temporal overlap and setting $t_0$.

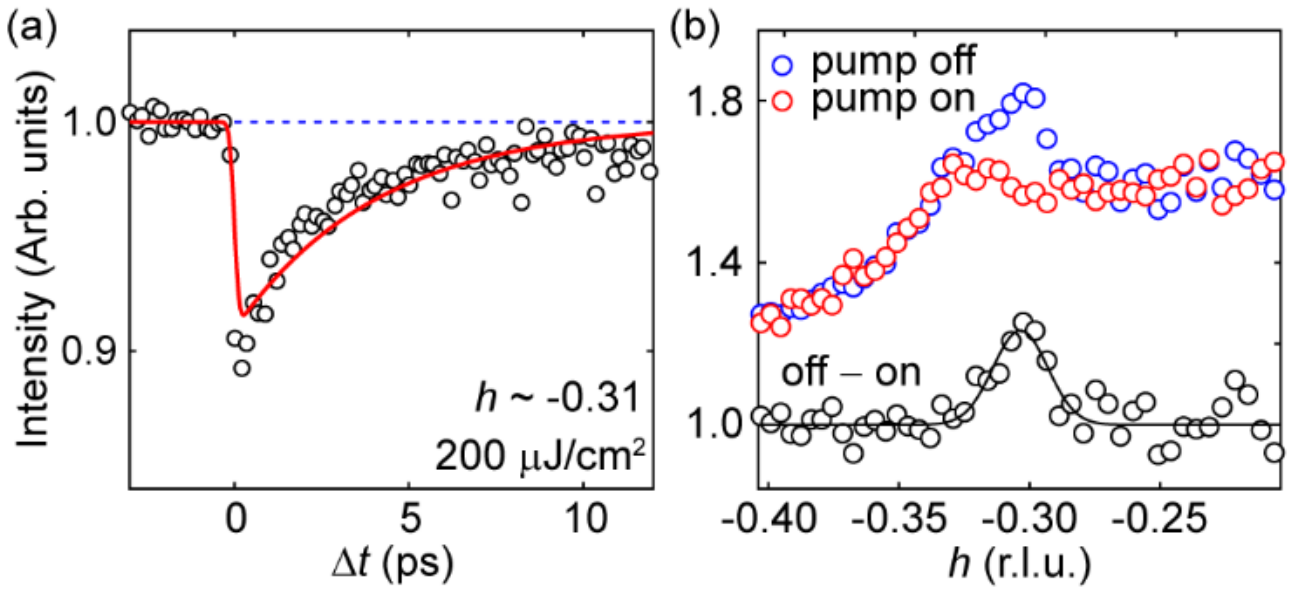

**FIG. 6**. (a) Delay time scan of the CDW intensity with a fluence of 200 μJ/cm$^2$ at $T$ = 70 K. The data are fitted by the convolution of a Gaussian with exponential decay. (b) The red and blue colored $h$ profiles are the CDW order with the pump at $\Delta t$ = 0.2 ps and without the pump, respectively. The black circles are the differences with a vertical shift. They are fitted by a single Gaussian curve.

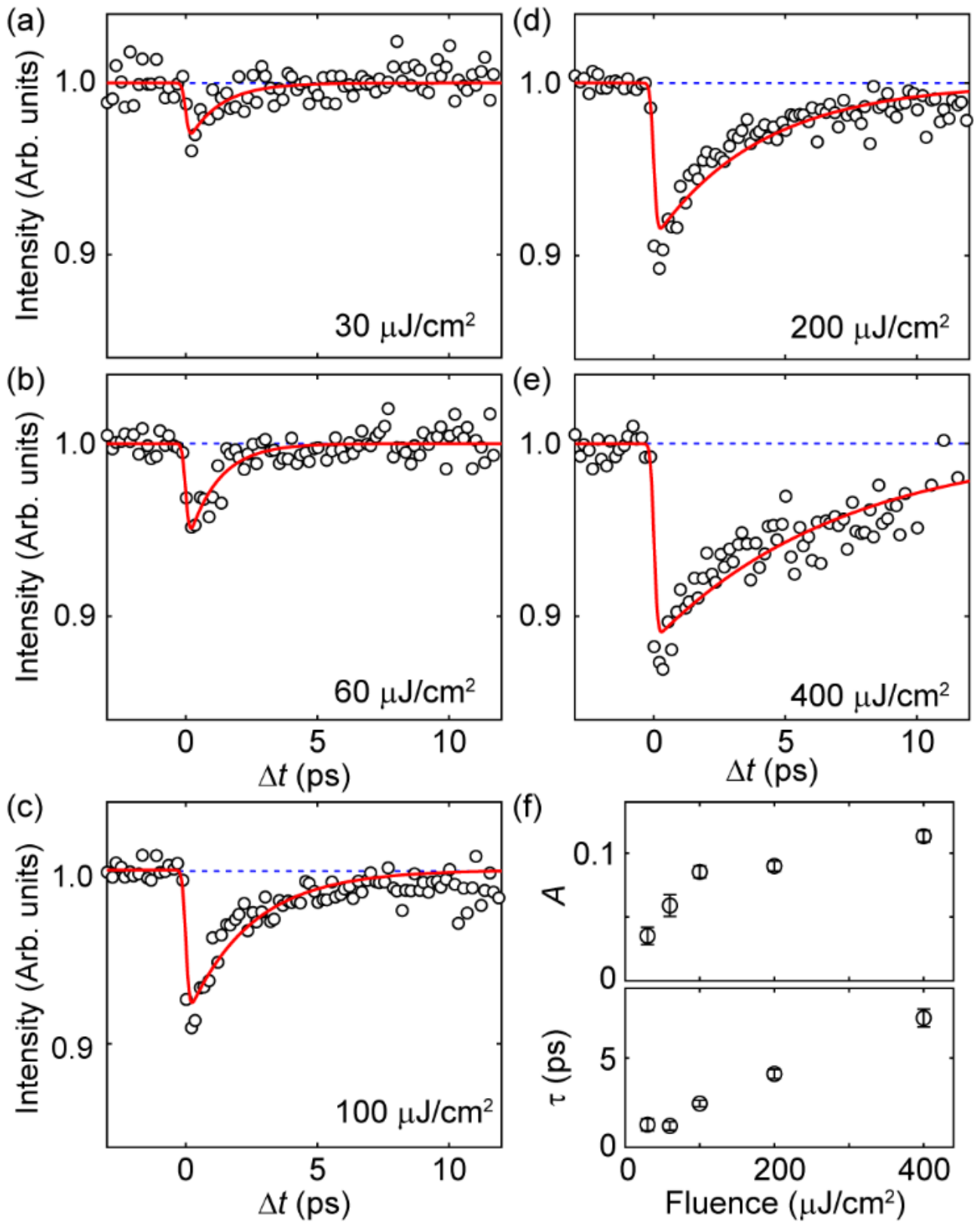

**FIG. 7**. (a)–(e) Delay time scans of the CDW intensity with various optical pump fluences. The data are fitted by the convolution of a Gaussian with exponential decay, as described in the text (red solid line). The blue dashed lines are base lines from $\Delta t < 0$. (f) Fits for amplitude ($A$) and decay time ($\tau$) using fitting function Eq. (2) are presented in the upper and lower panels, respectively. The error bars represent 1 standard deviation.

As shown in Fig. 6(a), we performed tr-RSXS with a fluence of 200 μJ/cm$^2$ to explore the temporal dynamics of the CDW order of YBCO. At $\Delta t \approx 0.2$ ps, the melting behavior of the CDW order is clearly shown. To confirm the melting behavior in momentum space, we ran an $h$ profile scan on the CDW order at $\Delta t = 0.2$ ps and compared it with the CDW order without an optical pump. On the one hand, the peak profiles around $h \approx -0.33$, with and without the optical pump, which corresponds to the OO tail, are similar, implying the OO is robust in this range of pump fluence. On the other hand, there is a clear difference around $h \approx -0.31$. We also highlight the differences between the $h$ profiles with the pump on and off [black circles in Fig. 6(b)], i.e., photo-induced melted intensity, which is fitted well by a single Gaussian curve at $h \approx -0.31$. These results indicate that the decrease in the scattering intensity is mainly due to the CDW melting behavior.

To learn more about the CDW dynamics, we investigated the temporal dynamics with various optical pump fluences (30, 60, 100, 200, and 400 μJ/cm$^2$). As shown in Fig. 7, the CDW melting tends to increase with an increase of the fluence. In addition, after melting the CDW order within a few hundred fs, the recovery of the CDW intensities tends to be slower as the fluence increases. For a better understanding, the temporal intensity change was fitted by a convolution of a Gaussian with exponential decay:

$$\Delta I = A \int_0^{\infty} dt' \exp\left(-\frac{\Delta t}{\tau}\right) \frac{1}{\sigma\sqrt{2\pi}} \exp\left(-\frac{(\Delta t - t')^2}{2\sigma^2}\right) \qquad (2)$$

where $A$ is the amplitude, $\tau$ the decay time, and $\sigma$ is the width of the Gaussian distribution. With $\sigma$ fixed at 80 fs, we estimated the amplitude and decay time for each fluence, as presented in Fig. 7(f). On increasing the fluence, the estimated decay time also increases, from ~1 ps (at 30 μJ/cm$^2$) to ~7 ps (at 400 μJ/cm$^2$), clearly showing slower recovery. The slower recovery for a higher fluence is understood as being due to a laser-induced thermal effect. Although it is too early to draw scientific conclusions, these findings may help in understanding the complex intertwining phenomena between the CDW order and the superconductivity of YBCO. Moreover, this experimental demonstration proves that the newly developed tr-RSXS system at PAL-XFEL is working properly. To help the readers to estimate the scanning time, we note that the photon number per pulse is about $4 \times 10^{10}$ in the current setting (930 eV and 200 μm exit slit width) and the delay scans in Fig. 7 took from 12 minutes (400 μJ/cm$^2$) to 37 minutes (200 μJ/cm$^2$) with 60 Hz X-ray repetition rate.

## V. OUTLOOK

In addition to the current work, we have several plans to upgrade the endstation and the beamline. First of all, we will install PERCIVAL, which is a dedicated CMOS 2D area detector for SX. This instrument is under development by an international collaboration of light sources.[47] This detector covers the $E_{ph}$ range of the SSS beamline well and the repetition rate is expected to be up to 300 Hz. High-throughput measurements based on shot-to-shot 2D images will be possible. Moreover, the background acquisition and wide-angle integration will be easier. PERCIVAL will be mounted on the same detector stage as, and is complementary to, the existing APD point detector. Due to increasing demands for a longer wavelength pump, e.g., a mid-infrared or THz laser,[48,49] we are planning to set up a THz generation system for the optical laser system in the near future. In the current plan, THz excitation will be injected to the side flange (30° from the X-ray path) and the coupling mirror will be installed inside the main chamber. Moreover, a phase shifter-based delay control scheme that allows longer delay times between the pump and probe pulses will be developed soon. Finally, we have proposed installing an elliptically polarized undulator in the SSS beamline as an afterburner configuration,[50] which can provide various polarization options, including circular and vertical linear, in addition to the current horizontal linear polarization. This would enable polarization-dependent tr-RSXS experiments, which may provide useful and accurate data on the complex ordering structure.

## VI. SUMMARY

A new RSXS endstation has been established at the SSS beamline of PAL-XFEL for ultrafast tr-RSXS experiments. This paper describes how the beamline and endstation were set up. The configuration of the hardware can be easily changed to suit the experimental requirements. The paper presents commissioning results for a high-$T_c$ cuprate superconductor. The RSXS endstation is open to the general user community and expected to host significant research into the temporal dynamics of strongly correlated systems and magnetic materials.

## ACKNOWLEDGMENTS

We appreciate the tremendous help and support from everyone at PAL-XFEL. We specially thank Jun-Sik Lee for insightful discussions and advice during the design phase of the tr-RSXS and during its commissioning. We acknowledge that the single-crystal YBCO was provided by Masaki Fujita. H.J. also appreciates the kind advices and comments from Byeong-Gyu Park, Yong-Sam Kim, Shunji Kishimoto, Joshua J. Turner, Alexander H. Reid, Georgi L. Dakovski, and Diling Zhu. The commissioning experiments were carried out at the SSS-RSXS endstation of PAL-XFEL funded by the Korea government (MIST). H.J. acknowledges the support by the National Research Foundation grant funded by the Korea government (MSIT) (No. 2019R1F1A1060295).

## AVAILABILITY OF DATA

The data that support the findings of this study are available from the corresponding author upon reasonable request.

## REFERENCES

[1]J. Stöhr, *NEXAFS Spectroscopy* (Springer-Verlag, Berlin and Heidelberg, 1992). https://doi.org/10.1007/978-3-662-02853-7

[2]S. W. Lovesey and S. P. Collins, X-Ray Scattering and Absorption from Magnetic Materials (Clarendon Press, Oxford, 1996).

[3]D. I. Khomskii and G. A. Sawatzky, Solid State Commun. **102**, 87 (1997). https://doi.org/10.1016/S0038-1098(96)00717-X

[4]E. Dagotto, Science **309**, 257 (2005). https://doi.org/10.1126/science.1107559

[5]C. Kao, J. B. Hastings, E. D. Johnson, D. P. Siddons, G. C. Smith, and G. A. Prinz, Phys. Rev. Lett. **65**, 373 (1990). https://doi.org/10.1103/PhysRevLett.65.373

[6]H. Wadati, "Resonant soft X-ray scattering studies of transition-metal oxides," in *Resonant X-Ray Scattering in Correlated Systems*, Springer Tracts in Modern Physics, edited by Y. Murakami and S. Ishihara, (Springer-Verlag, Berlin and Heidelberg, 2017), pp. 159–196. https://doi.org/10.1007/978-3-662-53227-0_5

[7]J. Fink, E. Schierle, E. Weschke, and J Geck, Rep. Prog. Phys. **76**, 056502 (2013). https://doi.org/10.1088/0034-4885/76/5/056502

[8]P. Emma, R. Akre, J. Arthur, R. Bionta, C. Bostedt, J. Bozek, A. Brachmann, P. Bucksbaum, R. Coffee, F.-J. Decker, Y. Ding, D. Dowell, S. Edstrom, A. Fisher, J. Frisch, S. Gilevich, J. Hastings, G. Hays, Ph. Hering, Z. Huang, R. Iverson, H. Loos, M. Messerschmidt, A. Miahnahri, S. Moeller, H.-D. Nuhn, G. Pile, D. Ratner, J. Rzepiela, D. Schultz, T. Smith, P. Stefan, H. Tompkins, J. Turner, J.Welch,W. White, J.Wu, G. Yocky, and J. Galayda, Nat. Photonics **4**, 641 (2010). https://doi.org/10.1038/nphoton.2010.176

[9]C. Bostedt, S. Boutet, D. M. Fritz, Z. Huang, H. J. Lee, H. T. Lemke, A. Robert, W. F. Schlotter, J. J. Turner, and G. J. Williams, Rev. Mod. Phys. **88**, 015007 (2016). https://doi.org/10.1103/RevModPhys.88.015007

[10]M. Yabashi, H. Tanaka, and T. Ishikawa, J. Synchrotron Radiat. **22**, 477 (2015). https://doi.org/10.1107/S1600577515004658

[11]I. S. Ko, H.-S. Kang, H. Heo, C. Kim, G. Kim, C.-K. Min, H. Yang, S. Y. Baek, H.-J. Choi, G. Mun, B. R. Park, Y. J. Suh, D. C. Shin, J. Hu, J. Hong, S. Jung, S.-H. Kim, K. Kim, D. Na, S. S. Park, Y. J. Park, Y. G. Jung, S. H. Jeong, H. G. Lee, S. Lee, S. Lee, B. Oh, H. S. Suh, J.-H. Han, M. H. Kim, N.-S. Jung, Y.-C. Kim, M.-S. Lee, B.-H. Lee, C.-W. Sung, I.-S. Mok, J.-M. Yang, Y. W. Parc, W.-W. Lee, C.-S. Lee, H. Shin, J. H. Kim, Y. Kim, J. H. Lee, S.-Y. Park, J. Kim, J. Park, I. Eom, S. Rah, S. Kim, K. H. Nam, J. Park, J. Park, S. Kim, S. Kwon, R. An, S. H. Park, K. S. Kim, H. Hyun, S. N. Kim, S. Kim, C.-J. Yu, B.-S. Kim, T.-H. Kang, K.-W. Kim, S.-H. Kim, H.-S. Lee, H.-S. Lee, K.-H. Park, T.-Y. Koo, D.-E. Kim, and K. B. Lee, Appl. Sci. **7**, 479 (2017). https://doi.org/10.3390/app7050479

[12]H.-S. Kang, C.-K. Min, H. Heo, C. Kim, H. Yang, G. Kim, I. Nam, S. Y. Baek, H.-J. Choi, G. Mun, B. R. Park, Y. J. Suh, D. C. Shin, J. Hu, J. Hong, S. Jung, S.-H. Kim, K. H. Kim, D. Na, S. S. Park, Y. J. Park, J.-H. Han, Y. G. Jung, S. H. Jeong, H. G. Lee, S. Lee, S. Lee, W.-W. Lee, B. Oh, H. S. Suh, Y. W. Parc, S.-J. Park, M. H. Kim, N.-S. Jung, Y.-C. Kim, M.-S. Lee, B.-H. Lee, C.-W. Sung, I.-S. Mok, J.-M. Yang, C.-S. Lee, H. Shin, J. H. Kim, Y. Kim, J. H. Lee, S.-Y. Park, J. Kim, J. Park, I. Eom, S. Rah, S. Kim, K. H. Nam, J. Park, J. Park, S. Kim, S. Kwon, S. H. Park, K. S. Kim, H. Hyun, S. N. Kim, S. Kim, S.-m. Hwang, M. J. Kim, C.-y. Lim, C.-J. Yu, B.-S. Kim,

T.-H. Kang, K.-W. Kim, S.-H. Kim, H.-S. Lee, H.-S. Lee, K.-H. Park, T.-Y. Koo, D.-E. Kim, and I. S. Ko, Nat. Photonics **11**, 708 (2017). https://doi.org/10.1038/s41566-017-0029-8

[13]H. Sinn, M. Dommach, X. Dong, D. La Civita, L. Samoylova, R. Villanueva, and F. Yang, Technical Design Report XFEL.EU TR-2012-006, 2012. https://doi.org/10.3204/XFEL.EU/TR-2012-006

[14]R. Abela, P. Beaud, J. A. van Bokhoven, M. Chergui, T. Feurer, J. Haase, G. Ingold, S. L. Johnson, G. Knopp, H. Lemke, C. J. Milne, B. Pedrini, and P. Radi, Struct. Dyn. **4**, 061602 (2017). https://doi.org/10.1063/1.4997222

[15]W. F. Schlotter, J. J. Turner, M. Rowen, P. Heimann, M. Holmes, O. Krupin, M. Messerschmidt, S. Moeller, J. Krzywinski, R. Soufli, M. Fernández-Perea, N. Kelez, S. Lee, R. Coffee, G. Hays, M. Beye, N. Gerken, F. Sorgenfrei, S. Hau-Riege, L. Juha, J. Chalupsky, V. Hajkova, A. P. Mancuso, A. Singer, O. Yefanov, I. A. Vartanyants, G. Cadenazzi, B. Abbey, K. A. Nugent, H. Sinn, J. Lüning, S. Schaffert, S. Eisebitt, W.-S. Lee, A. Scherz, A. R. Nilsson, and W. Wurth, Rev. Sci. Instrum. **83**, 043107 (2012). https://doi.org/10.1063/1.3698294

[16]G. L. Dakovski, P. Heimann, M. Holmes, O. Krupin, M. P. Minitti, A. Mitra, S. Moeller, M. Rowen, W. F. Schlotter, and J. J. Turner, J. Synchrotron Radiat. **22**, 498 (2015). https://doi.org/10.1107/S160057751500301X

[17]J. J. Turner, G. L. Dakovski, M. C. Hoffmann, H. Y. Hwang, A. Zarem, W. F. Schlotter, S. Moeller, M. P. Minitti, U. Staub, S. Johnson, A. Mitra, M. Swiggers, P. Noonan, G. I. Curiel, and M. Holmes, J. Synchrotron Radiat. **22**, 621 (2015). https://doi.org/10.1107/S1600577515005998

[18]D. J. Higley, K. Hirsch, G. L. Dakovski, E. Jal, E. Yuan, T. Liu, A. A. Lutman, J. P. MacArthur, E. Arenholz, Z. Chen, G. Coslovich, P. Denes, P. W. Granitzka, P. Hart, M. C. Hoffmann, J. Joseph, L. L. Guyader, A. Mitra, S. Moeller, H. Ohldag, M. Seaberg, P. Shafer, J. Stöhr, A. Tsukamoto, H.-D. Nuhn, A. H. Reid, H. A. Dürr, and W. F. Schlotter, Rev. Sci. Instrum. **87**, 033110 (2016). https://doi.org/10.1063/1.4944410

[19]D. Doering, Y.-D. Chuang, N. Andresen, K. Chow, D. Contarato, C. Cummings, E. Domning, J. Joseph, J. S. Pepper, B. Smith, G. Zizka, C. Ford, W. S. Lee, M. Weaver, L. Patthey, J. Weizeorick,

Z. Hussain, and P. Denes, Rev. Sci. Instrum. **82**, 073303 (2011). https://doi.org/10.1063/1.3609862

[20]M. Först, R. I. Tobey, S. Wall, H. Bromberger, V. Khanna, A. L. Cavalieri, Y.-D. Chuang, W. S. Lee, R. Moore, W. F. Schlotter, J. J. Turner, O. Krupin, M. Trigo, H. Zheng, J. F. Mitchell, S. S. Dhesi, J. P. Hill, and A. Cavalleri, Phys. Rev. B **84**, 241104 (2011). https://doi.org/10.1103/PhysRevB.84.241104

[21]W. S. Lee, Y. D. Chuang, R. G. Moore, Y. Zhu, L. Patthey, M. Trigo, D. H. Lu, P. S. Kirchmann, O. Krupin, M. Yi, M. Langner, N. Huse, J. S. Robinson, Y. Chen, S. Y. Zhou, G. Coslovich, B. Huber, D. A. Reis, R. A. Kaindl, R. W. Schoenlein, D. Doering, P. Denes, W. F. Schlotter, J. J. Turner, S. L. Johnson, M. Först, T. Sasagawa, Y. F. Kung, A. P. Sorini, A. F. Kemper, B. Moritz, T. P. Devereaux, D.-H. Lee, Z. X. Shen, and Z. Hussain, Nat. Commun. **3**, 838 (2012). https://doi.org/10.1038/ncomms1837

[22]T. Kubacka, J. A. Johnson, M. C. Hoffmann, C. Vicario, S. de Jong, P. Beaud, S. Grübel, S.-W. Huang, L. Huber, L. Patthey, Y.-D. Chuang, J. J. Turner, G. L. Dakovski, W.-S. Lee, M. P. Minitti, W. Schlotter, R. G. Moore, C. P. Hauri, S. M. Koohpayeh, V. Scagnoli, G. Ingold, S. L. Johnson, and U. Staub, Science **343**, 1333 (2014). https://doi.org/10.1126/science.1242862

[23]M. Först, R. I. Tobey, H. Bromberger, S. B. Wilkins, V. Khanna, A. D. Caviglia, Y.-D. Chuang, W. S. Lee, W. F. Schlotter, J. J. Turner, M. P. Minitti, O. Krupin, Z. J. Xu, J. S. Wen, G. D. Gu, S. S. Dhesi, A. Cavalleri, and J. P. Hill, Phys. Rev. Lett. **112**, 157002 (2014). https://doi.org/10.1103/PhysRevLett.112.157002

[24]M. Mitrano, S. Lee, A. A. Husain, L. Delacretaz, M. Zhu, G. de la Peña Munoz, S. X.-L. Sun, Y. I. Joe, A. H. Reid, S. F. Wandel, G. Coslovich, W. Schlotter, T. van Driel, J. Schneeloch, G. D. Gu, S. Hartnoll, N. Goldenfeld, and P. Abbamonte, Sci. Adv. **5**, eaax3346 (2019). https://doi.org/10.1126/sciadv.aax3346

[25]S. Wandel, F. Boschini, E. H. da Silva Neto, L. Shen, M. X. Na, S. Zohar, Y. Wang, G. B. Welch, M. H. Seaberg, J. D. Koralek, G. L. Dakovski, W. Hettel, M.-F. Lin, S. P. Moeller, W. F. Schlotter, A.H. Reid, M. P. Minitti, T. Boyle, F. He, R. Sutarto, R. Liang, D. Bonn, W. Hardy, R. A. Kaindl, D. G. Hawthorn, J.-S. Lee, A. F. Kemper, A. Damascelli, C. Giannetti, J. J. Turner, and G. Coslovich, arXiv:2002.02726 (preprint) (2020). https://arxiv.org/abs/2003.04224

[26]S. H. Park, M. S. Kim, C.-K. Min, I. E. Eom, I. Nam, H.-S. Lee, H.-S. Kang, H. D. Kim, H. Jang, J. Park, T.-Y. Koo, and S. Kwon, Rev. Sci. Instrum. **89**, 055105 (2018). https://doi.org/10.1063/1.5023557

[27]S. H. Park, J. Yoon, C. Kim, C. Hwang, D.-H. Kim, S.-H. Lee, and S. Kwon, J. Synchrotron Radiat. **26**, 1031 (2019). https://doi.org/10.1107/S1600577519004272

[28]A. S. M. Ismail, Y. Uemura, S. H. Park, S. Kwon, M. Kim, H. Elnaggar, F. Frati, Y. Niwa, H. Wadati, Y. Hirata, Y. Zhang, K. Yamagami, S. Yamamoto, I. Matsuda, U. Halisdemir, G. Koster, B. M. Weckhuysen, and F. M. F. de Groot, Phys. Chem. Chem. Phys., **22**, 2685 (2020). https://doi.org/10.1039/C9CP03374B

[29]D.-E. Kim, Y.-G. Jung, W.-W. Lee, K.-H. Park, S.-B. Lee, B.-G. Oh, S.-H. Jeong, H.-G. Lee, H.-S. Suh, H.-S. Kang, I.-S. Ko, and J. Pflueger, J. Korean Phys. Soc. **71**, 744 (2017). https://doi.org/10.3938/jkps.71.744

[30]Z. Huang and K.-J. Kim, Phys. Rev. Spec. Top. Accel. Beams **10**, 034801 (2007). https://doi.org/10.1103/PhysRevSTAB.10.034801

[31]K. Tiedtke, J. Feldhaus, U. Hahn, U. Jastrow, T. Nunez, T. Tschentscher, S. V. Bobashev, A. A. Sorokin, J. B. Hastings, S. Moller, L. Cibik, A. Gottwald, A. Hoehl, U. Kroth, M. Krumrey, H. Schoppe, G. Ulm, and M. Richter, J. Appl. Phys. **103**, 094511 (2008). https://doi.org/10.1063/1.2913328

[32]S. Moeller, G. Brown, G. Dakovski, B. Hill, M. Holmes, J. Loos, R. Maida, E. Paiser, W. Schlotter, J. J. Turner, A. Wallace, U. Jastrow, S. Kreis, A. A. Sorokin, and K. Tiedtke, J. Synchrotron Radiat. **22**, 606 (2015). https://doi.org/10.1107/S1600577515006098

[33]M. Kim, C.-K. Min, and I. Eom, J. Synchrotron Radiat. **26**, 868 (2019). https://doi.org/10.1107/S1600577519003515

[34]Experimental Physics and Industrial Control System (EPICS), http://www.aps.anl.gov/epics

[35]L. R. Dalesio, M. R. Kraimer, and A. J. Kozubal. (Proc. of ICALEPCS, KEK, Tsukuba, Japan, 1991), pp. 278. http://www.aps.anl.gov/epics/EpicsDocumentation/EpicsGeneral/EPICS_Architecture.pdf

[36]A. Sanchez-Gonzalez, A. S. Johnson, A. Fitzpatrick, C. D. M. Hutchison, C. Fare, V. Cordon-Preciado, G. Dorlhiac, J. L. Ferreira, R. M. Morgan, J. P. Marangos, S. Owada, T. Nakane, R. Tanaka, K. Tono, S. Iwata, and J. J. van Thor, J. Appl. Phys. **122**, 203105 (2017). https://doi.org/10.1063/1.5012749

[37]T. Sato, J. M. Glownia, M. R. Ware, M. Chollet, S. Nelsona, and D. Zhu, J. Synchrotron Radiat. **26**, 647 (2019). https://doi.org/10.1107/S1600577519002248

[38]M. Harmand, R. Coffee, M. R. Bionta, M. Chollet, D. French, D. Zhu, D. M. Fritz, H. T. Lemke, N. Medvedev, B. Ziaja, S. Toleikis, and M. Cammarata, Nat. Photonics **7**, 215 (2013). https://doi.org/10.1038/nphoton.2013.11

[39]R. Mankowsky, A. Subedi, M. Först, S. O. Mariager, M. Chollet, H. T. Lemke, J. S. Robinson, J. M. Glownia, M. P. Minitti, A. Frano, M. Fechner, N. A. Spaldin, T. Loew, B. Keimer, A. Georges, and A. Cavalleri, Nature **516**, 71 (2014). https://doi.org/10.1038/nature13875

[40]S. Gerber, H. Jang, H. Nojiri, S. Matsuzawa, H. Yasumura, D. A. Bonn, R. Liang, W. N. Hardy, Z. Islam, A. Mehta, S. Song, M. Sikorski, D. Stefanescu, Y. Feng, S. A. Kivelson, T. P. Devereaux, Z.-X. Shen, C.-C. Kao, W.-S. Lee, D. Zhu, and J.-S. Lee, Science **350**, 949 (2015). https://doi.org/10.1126/science.aac6257

[41]H. Jang, W.-S. Lee, H. Nojiri, S. Matsuzawa, H. Yasumura, L. Nie, A. V. Maharaj, S. Gerber, Y.-J. Liu, A. Mehta, D. A. Bonn, R. Liang, W. N. Hardy, C. A. Burns, Z. Islam, S. Song, J. Hastings, T. P. Devereaux, Z.-X. Shen, S. A. Kivelson, C.-C. Kao, D. Zhu, and J.-S. Lee, PNAS **113**, 14645 (2016). https://doi.org/10.1073/pnas.1612849113

[42]T. Naito, T. Nishizaki, Y. Watanabe, and N. Kobayashi, "Preparation and magnetic properties of untwinned $YBa_2Cu_3O_y$ single crystals," in *Advances in Superconductivity*, edited by S. Nakajima and M. Murakami (Springer-Verlag, Tokyo, 1997), Vol. IX, p. 601. https://www.springer.com/gp/book/9784431701989

[43]A. J. Achkar, R. Sutarto, X. Mao, F. He, A. Frano, S. Blanco-Canosa, M. Le Tacon, G. Ghiringhelli, L. Braicovich, M. Minola, M. Moretti Sala, C. Mazzoli, Ruixing Liang, D. A. Bonn, W. N. Hardy, B. Keimer, G. A. Sawatzky, and D. G. Hawthorn, Phys. Rev. Lett. **109**, 167001 (2012). https://doi.org/10.1103/PhysRevLett.109.167001

[44]H. Huang, H. Jang, M. Fujita, T. Nishizaki, Y. Lin, J. Wang, J. Ying, J. S. Smith, C. Kenney-Benson, G. Shen, W. L. Mao, C.-C. Kao, Y.-J. Liu, and J.-S. Lee, Phys. Rev. B **97**, 174508 (2018). https://doi.org/10.1103/PhysRevB.97.174508

[45]R. Comin and A. Damascelli, Annu. Rev. Condens. Matter Phys. **7**, 369 (2016). https://doi.org/10.1146/annurev-conmatphys-031115-011401

[46]M. v. Zimmermann, J. R. Schneider, T. Frello, N. H. Andersen, J. Madsen, M. Käll, H. F. Poulsen, R. Liang, P. Dosanjh, and W. N. Hardy, Phys. Rev. B **68**, 104515 (2003). https://doi.org/10.1103/PhysRevB.68.104515

[47]C. B. Wunderer, J. Correa, A. Marras, S. Aplin, B. Boitrelle, P. Goettlicher, F. Krivan, M. Kuhn, S. Lange, M. Niemann, F. Okrent, I. Shevyakov, M. Zimmer, N. Guerrini, B. Marsh, I. Sedgwick, G. Cautero, D. Giuressi, I. Gregori, G. Pinaroli, R. Menk, L. Stebel, A. Greer, T. Nicholls, U. K. Pedersen, N. Tartoni, H. Hyun, K. Kim, S. Rah, and H. Graafsma, J. Instrum. **14**, C01006 (2019). https://doi.org/10.1088/1748-0221/14/01/C01006

[48]T. Kampfrath, K. Tanaka, and K. A. Nelson, Nat. Photonics **7**, 680 (2013). https://doi.org/10.1038/nphoton.2013.184

[49]R. Mankowsky, M. Först, and A. Cavalleri, Rep. Prog. Phys. **79**, 064503 (2016). https://doi.org/10.1088/0034-4885/79/6/064503

[50]E. A. Schneidmiller and M. V. Yurkov, Phys. Rev. Spec. Top. Accel. Beams **16**, 110702 (2013). https://doi.org/10.1103/PhysRevSTAB.16.110702